\documentclass[12pt]{article}%
\usepackage[applemac]{inputenc}
\usepackage{amsmath,amssymb}
\usepackage{amsfonts}
\usepackage{amsmath,amssymb}
\usepackage[applemac]{inputenc}
\usepackage{color}
\usepackage{color, colortbl}
\usepackage{amsmath}
\usepackage{amssymb}
\usepackage{bbm}
\usepackage{graphicx}
\usepackage{tikz}
\usepackage{geometry}
\usetikzlibrary{arrows}
\usepackage{amsfonts}
\usepackage{amsmath,amssymb}
\usepackage[applemac]{inputenc}
\usepackage{color}
\usepackage{amsmath}
\usepackage{amssymb}
\usepackage{graphicx}
\usepackage{tikz}
\usepackage{bbm}
\usepackage{mathtools}
\usepackage{amssymb}
\usepackage{comment}
\usepackage{graphicx}
\usepackage{gensymb}
\usepackage{amsthm}
\usepackage{setspace}
\usepackage{graphicx}
\usepackage{subcaption}
\usepackage{mwe}
\usepackage{hyperref}
\usepackage{algorithm}
\usepackage{algorithmicx}
\usepackage{algpseudocode}

\newtheorem{theorem}{Theorem}[section]

\newtheorem{assumption}[theorem]{Assumption}

\newtheorem{remark}[theorem]{Remark}

\newcommand{\cA}{\mathcal{A}}
\newcommand{\cF}{\mathcal{F}}

\newcommand{\bF}{\mathbb{F}}
\newcommand{\bV}{\mathbb{V}}
\newcommand{\bP}{\mathbb{P}}
\newcommand{\bR}{\mathbb{R}}

\newcommand{\Nt}{\widetilde{N}}

\begin{document}
\title{A Deep Learning Approach to Renewable Capacity Installation under Jump Uncertainty}

\author{
Nacira Agram\textsuperscript{1,3} \and Fred Espen Benth\textsuperscript{2} \and Giulia Pucci\textsuperscript{1}   \and Jan Rems\textsuperscript{3}
}
\date{\today}
\maketitle

\begin{abstract}
We study a stochastic model for the installation of renewable energy capacity under demand uncertainty and jump driven dynamics. The system is governed by a multidimensional Ornstein-Uhlenbeck (OU) process driven by a subordinator, capturing abrupt variations in renewable generation and electricity load. Installation decisions are modeled through control actions that increase capacity in response to environmental and economic conditions.

We consider two distinct solution approaches. First, we implement a structured threshold based control rule, where capacity is increased proportionally when the stochastic capacity factor falls below a fixed level. This formulation leads to a nonlinear partial integro-differential equation (PIDE), which we solve by reformulating it as a backward stochastic differential equation with jumps. We extend the DBDP solver in \cite{hure2020deep} to the pure jump setting, employing a dual neural network architecture to approximate both the value function and the jump sensitivity.

Second, we propose a fully data driven deep control algorithm that directly learns the optimal feedback policy by minimizing the expected cost functional using neural networks. This approach avoids assumptions on the form of the control rule and enables adaptive interventions based on the evolving system state.

Numerical experiments highlight the strengths of both methods. While the threshold based BSDE approach offers interpretability and tractability, the deep control strategy achieves improved performance through flexibility in capacity allocation. Together, these tools provide a robust framework for decision support in long term renewable energy expansion under uncertainty.
\end{abstract}

\footnotetext[1]{Department of Mathematics, KTH Royal Institute of Technology, 100 44 Stockholm, Sweden.
Email: nacira@kth.se, pucci@kth.se. Supported by the Swedish Research Council grant (2020-04697).}

\footnotetext[2]{Department of Data Science and Analytics, 
BI Norwegian Business School, N-0484 Oslo, Nor-
way. E-mail: fred.e.benth@bi.no}

\footnotetext[3]{Department of Mathematics, University of Ljubljana, Ljubljana, Slovenia.
Email: jan.rems@fmf.uni-lj.si. Supported by the Slovenian Research and Innovation Agency, research core funding No. P1-0448.}

 

\section{Introduction}
\label{sec:introduction}

The expansion of renewable energy sources is essential to achieving a sustainable and resilient energy system. However, the intermittent and uncertain nature of renewable generation, such as wind and solar power, presents significant challenges for planning and decision making. A central question for energy planners is how to determine the timing and magnitude of new capacity installations in a way that ensures demand is met reliably while minimizing economic and environmental costs.

In this paper, we study a stochastic control problem for renewable capacity installation under uncertainty. The environment is modeled using a multidimensional Ornstein-Uhlenbeck (OU) process driven by a subordinator, capturing abrupt changes in both renewable generation potential and electricity demand. The objective is to determine an optimal control strategy for expanding renewable capacity over time in response to these uncertain dynamics.

\subsubsection*{Method 1: DeepBSDE-ControlSelector}
Our first approach is a structured, rule based strategy in which renewable capacity for technology $i$ is increased when the stochastic capacity factor falls below a fixed threshold $A_i$. The installation at time $t$ is given by
\[
 \left(A_i - V_i(t)\right)^+ \Delta V_i(t),
\]
where \( V(t) \in [0,1]^d \) is the vector capacity factors and \( \Delta V_i(t) \) is the corresponding jump. This policy is simple and interpretable, but restricts flexibility.  
For a given set of thresholds \( A = (A_1,\dots,A_d) \), the associated value function satisfies a nonlinear partial integro-differential equation (PIDE). Using a Feynman-Kac representation, this PIDE can be reformulated as a backward stochastic differential equation (BSDE) with jumps.  
We solve this BSDE numerically using an extension of the backward deep BSDE algorithm in \cite{hure2020deep} to pure jump settings. The method proceeds sequentially in time, reusing network parameters across steps, improving training stability and efficiency.  
This approach contrasts with the global forward deep BSDE methods in \cite{han2018solving, wang2025deep, alasseur2024deep, andersson2025deep} which approximate $(Y,Z,U)$ via forward simulation and a single terminal loss minimization.  In the BSDE with jumps framework, the triplet $(Y,Z,U)$ denotes the solution of the equation, where $Y$ is the value process of the control problem, $Z$ is the integrand with respect to the Brownian motion (capturing the sensitivity of $Y$ to continuous noise), and $U$ is the integrand with respect to the compensated Poisson random measure (capturing the sensitivity of $Y$ to jump events).

\subsubsection*{Method 2: DeepControlSolver}
Our second approach removes the threshold structure entirely. Inspired by the global deep control framework in \cite{han-control}, we model the feedback control law directly as a neural network and optimize it via stochastic gradient descent to minimize the expected total cost. This \emph{DeepControlSolver} does not require a Feynman-Kac representation or a derivable control rule, and can adapt to complex, state dependent or time varying decisions. This approach was extended to jump processes and applied to quadratic hedging problems in paper \cite{agram2024deep} for applications to hedging and in \cite{AGRAM2025106100} to conditional McKean Vlasov with jumps. While less interpretable, it often achieves lower cost values in simulation and provides a valuable benchmark for evaluating structured policies. This aligns with the recent paper \cite{ji2025novel}, in which the authors associate a control problem to the Hamiltonian system and solve it directly. 
\vspace{1em}

In our numerical experiments, the \emph{DeepControlSolver} achieves lower costs (e.g., $0.21$ vs.\ $0.29$), with the learned policy concentrating all investments in a single intervention, in line with the optimal single intervention behavior documented in \cite{agram2025installation}.
However, cost minimization is not the only criterion for evaluating the two approaches. Although the \emph{DeepBSDE-ControlSelector} incurs higher costs, it offers greater interpretability by providing a transparent rule that clearly indicates what to install and when.
Furthermore, it prescribes multiple interventions over the time horizon, which is often more consistent with real world decision making, where financial and logistical constraints favor gradual rather than one-off investments. From this perspective, the observed performance gap can be viewed as the "price" society pays for choosing a more interpretable and practically implementable rule, while still achieving outcomes that remain relatively close to the best attainable solution. The code containing the implementation of both approaches can be found on a GitHub repository \url{https://github.com/giuliapucci98/DeepEnergyInstallations}.

\vspace{1em}
\noindent\textbf{Organization of the paper.} 
Section~\ref{sec:stochastic_model} introduces the stochastic system dynamics. Section~\ref{sec:control-model} presents the optimal threshold control problem, derives the associated PIDE, and reformulates it as a BSDE with jumps. Additionally, we present the specialised $d=1$ model which will be later solved numerically. Section~\ref{sec:DeepBSDE} introduces the deep BSDE solver with jump extensions. Section~\ref{sec:numerics} provides numerical results. Section~\ref{sec:deepcontrol} presents the DeepControlSolver. Section~\ref{sec:conclusion} concludes.

\section{Stochastic Model for Capacity Factors and Demand}
\label{sec:stochastic_model}

To support robust decision making in energy system planning, it is essential to model the underlying drivers of both renewable energy production and electricity demand using stochastic processes. The model must capture not only the temporal dynamics but also the impact of unpredictable external events such as climate shocks or regulatory interventions.

We model the joint dynamics of energy production potential and demand using a multidimensional OU process with L\' evy jumps. The state variable $H(t) \in \mathbb{R}_+^{d+1}$ represents both the stochastic drivers of the capacity factor for $d$ renewable technologies and the driver of electricity demand.

\subsection{OU Process with Jumps}
Let $T > 0$ and $(\Omega, \cF, \bF, \bP)$ be a filtered probability space with a complete, right-continuous filtration $\bF = (\cF_t)_{t \in [0,T]}$. Suppose this space supports a Poisson random measure $N(dt, dz)$ with Levy measure $\nu$, and define the compensated random measure:
\[
\Nt(dt, dz) \coloneqq N(dt, dz) - \nu(dz) dt.
\]
We assume a finite activity setting, i.e., $\lambda = \int \nu(dz) < \infty$.

Consider $k$ different renewable technologies (e.g., wind, solar, hydro), each deployable at multiple spatial locations. Let $d = n_1 + \dots + n_k$ be the total number of location technology pairs. We define a $(d+1)$-dimensional stochastic process $H(t)$ governed by the SDE:
\begin{equation*}
dH(t) = - \Xi H(t) dt + \Sigma dL(t), \quad H(0) = x_0 \in \mathbb{R}_+^{d+1},
\end{equation*}
where:
\begin{itemize}
\item $\Xi = \text{diag}(\xi_1, \dots, \xi_{d+1})$ with $\xi_i > 0$ ensures mean-reverting dynamics;
\item $\Sigma \in \mathbb{R}^{(d+1) \times (d+1)}$ is a lower triangular matrix encoding spatial and technological dependencies;
\item $L(t)$ is a $d+1$-dimensional subordinator (i.e., an increasing pure-jump L\'evy process with positive jumps) with Poisson random measure $N(t,dz)$ and L\'evy measure $\nu(dz)$.

\end{itemize}

Using the L\' evy-It\^o decomposition, the jump process can be expressed as:
\begin{equation*}
L(t) = \int_0^t \int_{\mathbb{R}_+^{d+1}} z N(ds, dz) = \int_0^t \int_{\mathbb{R}_+^{d+1}} z \widetilde{N}(ds, dz) + t \int_{\mathbb{R}_+^{d+1}} z \nu(dz),
\end{equation*}
leading to the simplified dynamics:
\begin{equation*}
dH(t) = (b - \Xi H(t)) dt + \Sigma \int_{\mathbb{R}_+^{d+1}} z \widetilde{N}(dt, dz),
\end{equation*}
where $b = \Sigma \int_{\mathbb{R}_+^{d+1}} z \nu(dz)$.

\subsection{Capacity Factor and Demand Processes}
The renewable capacity factor vector $V(t) \in [0,1]^d$ is defined by a nonlinear transformation of $H(t)$:
\begin{equation}
V(t) = \mathbf{1} - \exp(-S(t)H(t)),
\label{eq:capacity}
\end{equation}
where $S(t) \in \mathbb{R}^{d \times (d+1)}$ is a diagonal matrix with an appended zero column with positive, deterministic entries $s_i(t)$ controlling seasonal effects in the dynamics of the capacity factor. The exponential is applied component-wise.

\begin{remark}
The structure $V(t) = \mathbf{1}  - \exp(-S(t)H(t))$ ensures that the capacity factor increases with increasing $H$, capturing the idea that better environmental conditions (e.g., more sun or wind) yield higher production efficiency.
\end{remark}

Demand is modeled as a scalar stochastic process $D(t)$ driven by the last component of $H(t)$:
\begin{equation*}
D(t) = p(t) (H_{d+1}(t)),
\end{equation*}
where $p(t)$ is a positive, deterministic function capturing seasonal or economic trends.

\subsection{Motivation and Empirical Basis}
The proposed stochastic structure reflects observed stylized properties of energy systems:
\begin{itemize}
\item Capacity factors exhibit reversion to seasonal baselines, punctuated by sudden volatility from meteorological or policy shocks;
\item Demand evolves with long-run socioeconomic trends, but responds to sudden weather changes and other electricity-demanding economic activities;
\item The multiplicative form $V(t) = \mathbf{1} - \exp(-S(t)H(t))$ is consistent with empirical models used in renewable energy forecasting. In particular, the paper \cite{BCR} explains well the probabilistic properties like the distribution and correlations across time, technologies and locations, in the case of wind capacity factors. See also \cite{benth2018non}.
\end{itemize}

\vspace{1em}
\noindent
This stochastic foundation supports the formulation of a dynamic control problem where installation decisions depend on the evolution of $H(t)$, as developed in the next section.
In particular, the capacity factor process \( V(t) \) will drive installation decisions through threshold-based policies, while the demand process \( D(t) \) enters the cost functional as a penalty for supply shortfall. The resulting control problem aims to determine optimal capacity adjustments in response to the stochastic evolution of \( H(t) \).

\section{Stochastic Control Framework for Renewable Installation}
\label{sec:control-model}

We develop a stochastic control framework to support decision-making in renewable energy investment under environmental uncertainty. The aim is to determine when and how much capacity should be installed to ensure that future electricity demand is met with minimal environmental and economic costs. Our model captures both the dynamic behavior of renewable resource availability and the demand process, incorporating jump risks to reflect abrupt changes due to climatic or market shocks.

 The state of the system at time \( t \in [0,T] \) is given by the triplet
\[
(V(t), D(t), C_R(t)) \in [0,1]^d \times \mathbb{R}_+ \times \mathbb{R}_+^d,
\]
where:
\begin{itemize}
    \item \( V(t) \in [0,1]^d \) is the vector of capacity factors for each technology or region, derived as \( V(t) = \mathbf{1} - \exp(-S(t) H(t)) \), where \( H(t) \in \mathbb{R}_+^{d+1} \) is a multivariate OU process;
    \item \( D(t) \in \mathbb{R}_+ \) denotes aggregate electricity demand, modeled as \( D(t) = p(t) H_{d+1}(t) \);
    \item \( C_R(t) \in \mathbb{R}_+^d \) is the cumulative installed renewable capacity vector.
\end{itemize}

The total renewable production at time \( t \) is given by \( V(t)^\top C_R(t) \), and any deficit relative to demand \( D(t) \) is compensated by fossil generation.

 In this model, we explicitly model capacity installations as being triggered by jumps in the underlying environment.   That is, new renewable installations occur only at such jump times.  The decision rule is given by a control policy \( A \in \cA \), which will be specified below. The dynamics of the installed capacity component evolve according to:
\[
dC_R^A(t) = \int_{\mathbb{R}_+} \beta_C(V(t), A(t), z) \, \widetilde{N}(dt,dz),
\]
where \( \widetilde{N} \) is the compensated Poisson random measure and \( \beta_C: [0,1]^d \times \mathbb{R}_+^d \times \mathbb{R} \to \mathbb{R}_+^{d \times d +1} \) determines the amount of new capacity installed in response to a jump.

The set of admissible controls \( \mathcal{A} \) consists of processes \( A \) with values in $\bR^d$ satisfying:
\begin{itemize}
    \item[(A1)] Progressive measurability with respect to the filtration \( \bF\);
    \item[(A2)] Component wise bounds \( A_{\min} \leq A_i(t) \leq A_{\max} \), for positive constants \(  A_{\min} < A_{\max} \) for all $t \in [0,T]$ and all \( i = 1,\dots,d \).
\end{itemize}

Let \( \kappa \in \mathbb{R}_+^d \) be the vector of installation costs per unit. The objective is to minimize the total expected cost, composed of the penalty for unmet demand and the terminal investment cost. The cost functional under policy \( A \in \mathcal{A} \) is defined as:
\begin{equation*}
\begin{aligned}
J^A(t, v,d,c) = \; &\mathbb{E} \bigg[ 
   \int_t^T e^{-rs}\left(D(s) - V(s)^\top C^A_R(s)\right)^+ ds  \\
   &\qquad + \; \kappa^\top C_R^A(T) \,\bigg|\, V(t) = v, D(t) = d, C^A_R(t) = c \bigg] .
\end{aligned}
\end{equation*}
where $r \ge 0$ is the discount factor and  \( C^A_R(t) \) denotes the cumulative installed capacity under control policy \( A \in \cA \). 
We define the value function:
\[
\bV(t,v,d,c) := \inf_{A \in \mathcal{A}} J^A(t,v,d,c),
\]
where \(  (v,d,c) \in [0,1]^d \times \mathbb{R}_+ \times \mathbb{R}_+^d \). The associated infinitesimal generator \( \mathcal{L}^A \) applied to a smooth function \( \varphi(t,v,d,c) \) is:
\small
\begin{align*}
\mathcal{L}^A \varphi(t,v,d,c) 
&= b_V(v)^\top \nabla_{v} \varphi(t,v,d,c) 
   + b_D(d)^\top \, \partial_{d} \varphi(t,v,d,c) 
   + b_C(c)^\top \nabla_{c} \varphi(t,v,d,c) \\
&\quad + \int_{\mathbb{R}_+^{d+1}} \Big[ 
     \varphi\big(t, v + \beta_V(v,z),\, d + \beta_D(d,z),\, c + \beta_C(v, A(t), z)\big) 
     - \varphi(t,v,d,c) \\
&\qquad\qquad 
     - \nabla_{(v,d,c)} \varphi(t,v,d,c) \cdot 
       \big(\beta_V(v,z),\, \beta_D(d,z),\, \beta_C(v, A(t), z)\big) 
   \Big] \nu(dz),
\end{align*}
where \( b_V(v) \) and \( b_D(d) \) denote the drift terms of the components \( V \) and \( D \), respectively; \( \beta_V(v, z) \) and \( \beta_D(d, z) \) represent their corresponding jump amplitudes in response to a jump of size \( z \in \mathbb{R} \); and \( \nu \) is the $d+1-$dimensional joint L\'evy measure. Here, $\nabla_v$ denotes the gradient with respect to the vector variable $v$, 
 $\partial_d$ denotes the partial derivative with respect to the scalar variable $d$, 
and 
\[
\nabla_{(v,d,c)} \varphi(t,v,d,c) :=
\big( \nabla_v \varphi(t,v,d,c), \; \partial_d \varphi(t,v,d,c), \; \nabla_c \varphi(t,v,d,c) \big)
\]
denotes the gradient with respect to all state variables jointly.

\begin{theorem}[Verification Theorem, adapted from \cite{OS}]
Suppose the value function \( \bV(t,v,d,c) \in C^{1}([0,T] \times  [0,1]^d \times \mathbb{R}_+ \times \mathbb{R}_+^d ) \) solves the HJB equation:
\begin{equation}
\label{eq:HJB}
\partial_t \bV + \mathcal{L}^A \bV + e^{-rt}\left(d - v^\top c\right)^+ = 0, \quad \bV(T,v,d,c) = \kappa^\top c.
\end{equation}
Assume further that the control \( A^* \in \mathcal{A} \) attains the infimum in the Hamilton-Jacobi-Bellman (HJB) equation. Then \( A^* \) is optimal and \( \bV \) is the value function.
\end{theorem}

\begin{remark} Notice that we use the space $C^{1}$ and not the more standard $C^{1,2}$, since there is no diffusion term in our dynamics. \end{remark}

\subsection{Threshold-Based Installation Policy}

Consider the following threshold-based installation rule. Let
\[
A = (A_1, \dots, A_d) \subset \bR_+^d
\]
be a vector of fixed thresholds. At any jump time of the capacity factor process \( V(t) \), the installed capacity is adjusted by
\begin{equation}
\Delta C^A_{R,i}(t) = (A_i - V_i(t-))^+ \Delta V_i(t), \quad i = 1, \dots. d, \label{eq:threshold}
\end{equation}
 Notice that $V$ is driven by an OU dynamics and decays continuously as stated by \eqref{eq:capacity}. Additionally, jumps in the underlying process lead to jumps up in V, i.e.  the jump magnitude  \( \Delta V_i(t) = V_i(t) - V_i(t^-) \) is always positive in the model. According to \eqref{eq:threshold}, capacity is installed precisely at such upward jumps, but only when the capacity factor has deteriorated below the threshold $A$. 
\begin{remark}
It should be noted that the control policy in \eqref{eq:threshold} is not the only possible specification for guiding the installation of renewable energy capacities. Other options include, for example, setting a fixed number of decision dates per year or triggering installations based on simple indicators of volatility. Alternative rules, such as threshold-based controls activated by changes in demand, were also considered. Among the policies explored, the proposed strategy was found to deliver the most favorable performance.
\end{remark}

While the verification theorem provides a theoretical characterization of the value function via the nonlinear partial PIDE \eqref{eq:HJB}, solving this equation explicitly or even numerically is extremely challenging, especially in high dimensions. The presence of a nonlinearity, jump terms, and threshold-based control structures makes standard finite difference or grid-based approaches computationally demanding, if not infeasible.

Therefore, in the subsequent sections, we reformulate the problem using a BSDE with jumps whose solution coincides with the value function under appropriate assumptions. We then extend recent deep learning methods to the jump diffusion setting. Finally, we develop an integrated framework that jointly solves the BSDE and optimizes over the threshold control policy using neural networks.

To gain insight into the behavior of the system and the role of threshold-based installation policies, we now specialize the general framework to the case of a single renewable technology at a single location. This two-dimensional setting preserves the essential stochastic features of the full model including mean reversion, jump dynamics, and state dependent capacity installation while allowing for clearer analytical interpretation and numerical tractability.

\subsection{Specialized Case: Two-Dimensional Model ({$d=1$})}
\label{sec:2dim_model}

To illustrate the structure and implications of the general control framework, we now consider the case of a single renewable technology at a single site, i.e., \( d = 1 \). In this simplified setting, the underlying state process reduces to a two-dimensional OU system that models the stochastic behavior of renewable potential and energy demand.

Let \( H(t) = (H_1(t), H_2(t))^\top \in \mathbb{R}_+^2 \), where \( H_1(t) \) governs the renewable capacity potential (e.g., related to wind or solar irradiance), and \( H_2(t) \) captures the demand driving factor. The state evolves according to:
\begin{equation*}
\begin{aligned}
\begin{bmatrix}
dH_1(t) \\
dH_2(t)
\end{bmatrix}
&= 
- \begin{bmatrix}
\xi_1 & 0 \\
0 & \xi_2
\end{bmatrix}
\begin{bmatrix}
H_1(t) \\
H_2(t)
\end{bmatrix} dt 
+ 
\begin{bmatrix}
\sigma_{11} & \sigma_{12} \\
0 & \sigma_{22}
\end{bmatrix}
\begin{bmatrix}
dL_1(t) \\
dL_2(t)
\end{bmatrix},
\end{aligned}
\end{equation*}
with \( H_i(0) = h_0^i \in \mathbb{R}_+ \), and where \( L_1(t) \), \( L_2(t) \) are independent compound Poisson processes with intensities \( \lambda_1, \lambda_2 > 0 \), and exponentially distributed jumps of mean \( 1/m_1 \), \( 1/m_2 \), respectively. Each \( L_i(t) \) admits the L\' evy-It\^o decomposition:
\begin{equation*}
L_i(t) = \int_0^t \int_{\mathbb{R}_+} z \, \widetilde{N}_i(ds, dz) + \frac{\lambda_i t}{m_i}, \quad i = 1,2,
\end{equation*}
where \( \widetilde{N}_i \) denotes the compensated Poisson random measure.

The renewable capacity factor process \( V(t) \in [0,1] \) is defined through a seasonal nonlinear transformation:
\begin{equation}
V(t) = 1 - \exp(-s_1(t) H_1(t)),
\label{eq:v_1d}
\end{equation}
where \( s_1(t) > 0 \) models seasonal variability. Applying It\^o's formula for jump processes, the dynamics of \( V(t) \) are:
\begin{equation}
    \begin{aligned}
dV(t) =& \; b_V(t,V(t))dt + \int_{\mathbb{R}_+}\beta^1_V(t,V(t),z) \widetilde{N}_1(dt, dz) + \int_{\mathbb{R}_+}\beta^2_V(t,V(t),z) \widetilde{N}_2(dt, dz)  \\ =&(1 - V(t)) \Bigg[
\left((\xi_1 - s_1'(t)) \ln(1 - V(t))
+ \frac{\lambda_1 s_1(t) \sigma_{11}}{m_1 + s_1(t)\sigma_{11}}
+ \frac{\lambda_2 s_1(t) \sigma_{12}}{m_2 + s_1(t)\sigma_{12}} \right) dt \\
&+ \int_{\mathbb{R}_+} \left(1 - e^{-\sigma_{11} z}\right) \, \widetilde{N}_1(dt, dz)
+ \int_{\mathbb{R}_+} \left(1 - e^{-\sigma_{12} z}\right) \, \widetilde{N}_2(dt, dz)
\Bigg]. \label{eq:V_din}
    \end{aligned}
\end{equation}
The energy demand process is modeled as:
\begin{equation*}
D(t) = p(t) H_2(t),
\end{equation*}
where \( p(t) \geq 0 \) is a deterministic seasonal load factor. $D(t)$ has dynamics: 
\begin{equation}
    \begin{aligned}
dD(t) =& \; b_D(t,D(t))dt  + \int_{\mathbb{R}_+}\beta^2_D(t,D(t),z) \widetilde{N}_2(dt, dz)  \\ =& \left[ p'(t)D(t) + p(t) (\frac{\lambda_2 \sigma_{22}}{m_2} - \xi_2 D(t)) \right ]dt  + \int_{\mathbb{R}_+} p(t) \sigma_{22} z \, \widetilde{N}_2(dt, dz)
. \label{eq:D_din}
    \end{aligned}
\end{equation}
To explore policy evaluation in this simple case, we consider a scalar threshold level \( A \in \mathcal{A} \subset \bR_+ \) governing the installation response and define the following threshold based rule for capacity adjustment:
\begin{equation}
\label{eq:threshold-1d}
\Delta C^A_R(t) = (A - V(t))^+ \Delta V(t),
\end{equation}
where \( C^A_R(t) \) denotes the cumulative installed capacity under control policy \( A \in \cA \subset \bR_+ \) and  \( \Delta V(t) = V(t) - V(t^-) \). 
Since $V(t)$ decays continuously between jumps (as in \eqref{eq:v_1d}, it may fall below the threshold $A$. At a jump time, $V$ receives a positive increment $\Delta V(t)$ from the underlying process $H$. The rule \eqref{eq:threshold-1d} installs capacity precisely at such upward jumps, but only if $V$ is below the threshold at that moment. The size of the installation is proportional to the jump magnitude, ensuring that capacity adjustments respond to upward shocks while respecting the threshold condition.

The corresponding expected cost under threshold policy \( A \) is:
\begin{equation}
\begin{aligned}
    \label{eq:cost-1d}
J^A(t,v,d,c) = \mathbb{E} \bigg[ \kappa C^A_R(T) + \int_t^T e^{-rs}\left(D(s) - V(s) C^A_R(s)\right)^+ ds \\ \bigg|\, V(t) = v, D(t) = d, C^A_R(t) = c \bigg],
\end{aligned}
\end{equation}
where \( \kappa > 0 \) is the unit installation cost and \( C^A_R(t) \) is the cumulative installed capacity. The optimization problem is:
\[
\begin{aligned}
\mathbb{V}(t, v, d, c) = \inf_{A \in \mathcal{A}} J^A(t,v,d,c).
\end{aligned}
\]
 The following PIDE characterizes the value function:

\begin{equation}
\label{eq:HJB-specialized}
\begin{aligned}
\partial_t \bV + \mathcal{L}^A \bV + e^{-rt} \left( d - v c \right)^+ = 0, \quad \bV(T, v, d, c) = \kappa c.
\end{aligned}
\end{equation}
The infinitesimal generator \( \mathcal{L}^A \) is given by:
\small
\[
\begin{aligned}
\mathcal{L}^A \bV(t,v,d,c) &= -b_V(t,v) \, \partial_{v} \bV - b_D(t,d)  \, \partial_{d} \bV \\
&\quad + \sum_{i=1}^2 \lambda_i \int_{\mathbb{R}_+} \Big[ \bV\left(t, v + \beta^i_V(t,v, z), d + \beta^i_D(t, d, z), c + \beta_C^i(v, A, z) \right) \\
&\qquad - \bV(t,v,d,c) - \beta^i_V(t,v,z) \, \partial_{v} \bV - \beta^i_D(t,d,z)  \, \partial_{d} \bV - \beta_C^i(v, A, z) \, \partial_c \bV \Big] f_i(z)\,dz.
\end{aligned}
\]
Here,
\[
\beta_C^i(v, A, z) := (A - v)^+ (1 - v) \left( 1 - e^{-\sigma_{1i} z} \right),
\]
and \( f_i(z) \) is the exponential probability density function for jump sizes with mean \( \frac{1}{m_i} \).

Equation \eqref{eq:HJB-specialized} incorporates the trade off between installation costs (terminal term), fossil backup penalties (integral term), and the random impact of capacity jumps. It defines a nonlinear, nonlocal equation that depends on the control variable \( A \), which enters both directly and through the jump impact function \( \delta \).

\noindent
\paragraph{Reformulation via BSDEs with Jumps.}
Solving the PIDE \eqref{eq:HJB-specialized} analytically is very hard due to the presence of nonlinearity, control-dependent jumps, and high-dimensional integration. Even numerical approaches such as finite differences are limited by the curse of dimensionality and poor convergence in the jump setting.

To overcome this, we reformulate the problem using a BSDE with jumps whose solution represents the value function \( \bV(t, V(t) ,D(t),C^A_R(t)) \). Letting \( (Y, U )\) denote the BSDE solution, the system reads:
\begin{equation}
\label{eq:bsde-specialized}
\begin{cases}
dY(t) = - e^{-rt}\left( D(t) - V(t) C^A_R(t) \right)^+ dt +\int_{\mathbb{R}} U_1(t,z) \, \widetilde{N}_1(dt,dz) + \int_{\mathbb{R}_+} U_2(t,z) \, \widetilde{N}_2(dt,dz), \\
Y(T) = \kappa C^A_R(T).
\end{cases}
\end{equation}
The processes \( U_i(t,z) \) capture the jump correction due to installation triggered by jumps in the forward process.\\

The BSDE \eqref{eq:bsde-specialized} offers a probabilistic characterization of the value function associated with the threshold based control problem. However, due to the presence of nonlinearities and discontinuities induced by jumps, it does not admit a closed form solution and remains challenging for classical numerical methods, particularly in higher dimensions.

To overcome these limitations, we build upon recent advances in deep learning methods for BSDEs  to construct an efficient approximation scheme for the value function and facilitate optimization over the threshold parameter \( A \). Specifically, we propose the \emph{	DeepBackwardBSDE-JumpSolver} (Algorithm \ref{algo:jumpFBSDE}), a neural network based architecture tailored to jump diffusion systems with discontinuous dynamics. This method extends existing deep BSDE techniques to accommodate threshold triggered control structures and discontinuous renewable supply processes.

In addition, we develop the \emph{DeepBSDE-ControlSelector} scheme (Algorithm \ref{algo:control-selection}), which systematically explores a grid of candidate threshold values to identify the optimal control policy \( A^* \). By coupling stochastic simulation of the forward process with backward resolution of the BSDE, our approach yields a data compatible and computationally scalable solution framework for renewable investment under uncertainty.

In contrast to fixed-threshold evaluation, we also propose a fully data driven method to directly learn the optimal control strategy. The resulting algorithm, called \emph{DeepControlSolver} (Algorithm \ref{algo:control}), estimates the feedback control policy via a neural network trained to minimize the cost functional over simulated state trajectories.

\section{Deep BSDEs with Jumps: Solving PIDEs via Probabilistic Representation}
\label{sec:DeepBSDE}

In this section, we extend the deep BSDE algorithm in \cite{hure2020deep} to handle pure jump dynamics. Our goal is to numerically solve nonlinear PIDEs that arise in jump diffusion models by employing feedforward neural networks and leveraging their backward stochastic representation via BSDEs with jumps.

\subsection{Neural Network Architecture for BSDE with jumps Approximation}
\label{sec:NN-intro}

We begin by briefly reviewing the structure of feedforward neural networks, which form the backbone of our deep BSDE solver. These networks provide a flexible, trainable framework for approximating nonlinear functions arising in high-dimensional stochastic control problems.

Let \( \mathcal{N}(x;\theta): \mathbb{R}^{d_0} \to \mathbb{R}^{d_1} \) denote a feedforward neural network parameterized by \( \theta \in \Theta_m \), where \( d_0 \) is the input dimension, \( d_1 \) is the output dimension, and \( \theta \) contains the collection of all trainable weights and biases. The network is constructed as a composition of affine transformations and nonlinear activation functions:
\begin{equation*}
\mathcal{N}(x;\theta) = (A_{L+1} \circ \rho \circ A_L \circ \rho \circ \dots \circ \rho \circ A_0)(x),
\end{equation*}
where \( L \in \mathbb{N}_0 \) denotes the number of hidden layers and \( \rho: \mathbb{R} \to \mathbb{R} \) is a fixed nonlinear activation function applied component-wise.

Each affine transformation \( A_\ell: \mathbb{R}^{d_\ell} \to \mathbb{R}^{d_{\ell+1}} \) for \( \ell = 0, \dots, L+1 \) is defined by:
\[
A_\ell(x) = W_\ell x + b_\ell,
\]
where \( W_\ell \in \mathbb{R}^{d_{\ell+1} \times d_\ell} \) is the weight matrix and \( b_\ell \in \mathbb{R}^{d_{\ell+1}} \) is the bias vector. For simplicity, we assume that all hidden layers have the same width \( m \), so that \( d_\ell = m \) for \( \ell = 1, \dots, L \), while \( d_0 \) and \( d_{L+1} \) represent the input and output dimensions, respectively.

The total number of trainable parameters in such a network is given by:
\[
N_m = m d_0 + m + L(m^2 + m) + d_1 m + d_1.
\]

We define the set of admissible parameter values by \( \Theta_m \subseteq \mathbb{R}^{N_m} \). When there are no structural constraints, we take \( \Theta_m = \mathbb{R}^{N_m} \).

Let \( \mathcal{NN}^{\rho}_{d_0,d_1,L,m}(\theta) \) denote the class of all feedforward neural networks of depth \( L \), width \( m \), activation \( \rho \), and parameters \( \theta \). We define the full neural network class with fixed structure as:
\[
\mathcal{NN}^{\rho}_{d_0,d_1,L} = \bigcup_{m \in \mathbb{N}} \mathcal{NN}^{\rho}_{d_0,d_1,L,m}(\theta).
\]

\begin{theorem}[Universal Approximation Theorem {\cite{hornik1989multilayer}}]
Let \( \rho \) be any continuous, nonconstant activation function. Then, for any finite Borel measure \( \nu \) on \( \mathbb{R}^{d_0} \), the class \( \mathcal{NN}^{\rho}_{d_0,d_1,L} \) is dense in \( L^2(\nu) \). That is, for any \( f \in L^2(\nu) \) and any \( \varepsilon > 0 \), there exists a neural network \( \mathcal{N} \in \mathcal{NN}^{\rho}_{d_0,d_1,L} \) such that
\[
\| f - \mathcal{N}(\cdot;\theta) \|_{L^2(\nu)} < \varepsilon.
\]
\end{theorem}

This result underpins the use of deep neural networks for approximating the unknown components in our BSDEs with jumps. In our setting, the neural network \( \mathcal{N} \) is used to approximate either the value process, the jump term, or the control feedback map, depending on the problem formulation.

\begin{figure}[h!]
\centering
\begin{tikzpicture}[scale=0.95, every node/.style={scale=0.95}]
    \tikzstyle{neuron}=[circle,draw,minimum size=18pt,inner sep=0pt]
    \tikzstyle{input neuron}=[neuron, fill=blue!30]
    \tikzstyle{output neuron}=[neuron, fill=red!30]
    \tikzstyle{hidden neuron}=[neuron, fill=gray!30]
    \tikzstyle{annot} = [text width=4em, text centered]

    \foreach \name / \y in {1,...,3}
        \node[input neuron] (I-\name) at (0,-\y) {};

    \foreach \name / \y in {1,...,4}
        \node[hidden neuron] (H1-\name) at (2,-\y + 0.5) {};
    \foreach \name / \y in {1,...,4}
        \node[hidden neuron] (H2-\name) at (4,-\y + 0.5) {};

    \foreach \name / \y in {1}
        \node[output neuron] (O-\name) at (6,-2) {};

    \foreach \source in {1,...,3}
        \foreach \dest in {1,...,4}
            \draw[->] (I-\source) -- (H1-\dest);

    \foreach \source in {1,...,4}
        \foreach \dest in {1,...,4}
            \draw[->] (H1-\source) -- (H2-\dest);

    \foreach \source in {1,...,4}
        \foreach \dest in {1}
            \draw[->] (H2-\source) -- (O-\dest);

    \node[annot, above of=I-1, node distance=1cm] {Input \\ $x \in \mathbb{R}^{d_0}$};
    \node[annot, above of=H1-1, node distance=1cm] {Hidden Layer 1};
    \node[annot, above of=H2-1, node distance=1cm] {Hidden Layer 2};
    \node[annot, above of=O-1, node distance=1cm] {Output \\ $\in \mathbb{R}^{d_1}$};

\end{tikzpicture}
\caption{Illustration of a feedforward neural network with two hidden layers, $d_0=3$ and $d_1=1$.}
\label{fig:nn-architecture}
\end{figure}
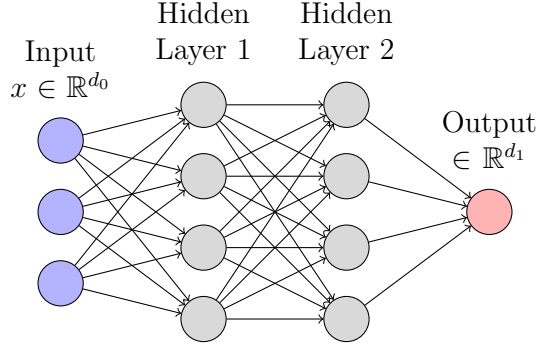

A schematic illustration of the network architecture is shown in Figure~\ref{fig:nn-architecture}.

\subsection{From PIDEs to BSDEs with Jumps}
We consider the pure jump stochastic differential equation (SDE):
\begin{equation}
X(s) = x + \int_t^s b(X(r))\,dr + \int_t^s \int_{\mathbb{R}^\ell} \beta(X(r), z)\, \widetilde{N}(dr, dz), \quad s \in [t, T],\ x \in \mathbb{R}^d,
\label{eq:forwardSDE}
\end{equation}
where $b : \mathbb{R}^d \to \mathbb{R}^d$ is the drift term and $\beta : \mathbb{R}^d \times \mathbb{R}^\ell \to \mathbb{R}^{d}$ is the jump coefficient, with $\widetilde{N}$ compensated Poisson random measure.

\begin{assumption}\label{ass:forward}
Let $b$ and $\beta$ satisfy the following:
\begin{enumerate}
    \item $b$ is Lipschitz continuous: \( |b(x) - b(x')| \leq K|x - x'| \) for some $K > 0$.
    \item $\beta$ satisfies:
    \begin{align*}
    \int_{|z| < 1} |\beta(x,z) - \beta(x',z)|^2 \nu(dz) &\leq K |x - x'|^2, \\
    \int_{|z| < 1} |\beta(x,z)|^2 \nu(dz) &\leq K(1 + |x|^2).
    \end{align*}
\end{enumerate}
\end{assumption}

We study the PIDE:
\begin{equation}
\label{eq:PIDE_general}
u(t,x) + \mathcal{L} u(t,x) + f(t,x,u(t,x)) = 0, \quad u(T,x) = g(x),
\end{equation}
where the integro-differential operator $\mathcal{L}$ is:
\small
\begin{align*}
\mathcal{L} u(t,x) = \langle b(x), \nabla_x u(t,x) \rangle + \int_{\mathbb{R}^d} \left[ u(t, x + \beta(x,z)) - u(t,x) - \langle \beta(x,z), \nabla_x u(t,x) \rangle \right] \nu(dz).
\end{align*}
This PIDE admits the following BSDE representation:
\begin{equation*}
Y(t) = g(X(T)) + \int_t^T f(r, X(r), Y(r)) dr - \int_t^T \int_{\mathbb{R}^d} U(r,z) \, \widetilde{N}(dr,dz).
\end{equation*}

\begin{assumption}\label{ass:backward}
The functions $f$ and $g$ satisfy:
\begin{enumerate}
    \item $f(t,x,y)$ is Lipschitz continuous in $x$ and $y$ uniformly in $t$.
    \item $g(x)$ is Lipschitz continuous in $x$.
\end{enumerate}
\end{assumption}

\begin{theorem}[Feynman-Kac Formula for BSDEs with jumps \cite{PIDE}]\label{thm:f-k}
Under Assumptions \ref{ass:forward} and \ref{ass:backward}, the unique solution $u$ of the PIDE \eqref{eq:PIDE_general} satisfies:
\[
Y^{t,x}(s) = u(s, X^{t,x}(s)), \quad U^{t,x}(s,z) = u(s, X^{t,x}(s) + \beta(X^{t,x}(s), z)) - u(s, X^{t,x}(s)).
\]
\end{theorem}
\subsection{Discretization and Neural Approximation of BSDEs with Jumps}

In this work, we extend the numerical resolution of BSDEs to the jump setting, building on the foundational backward in time approach for BSDEs without jumps introduced in \cite{hure2020deep}, and the theoretical framework of \cite{castro2022deep}, which incorporates jumps but remains purely analytical and without a numerical implementation. While \cite{castro2022deep} focuses on approximating only the integrand of the jump term, our method directly approximates the entire jump integral in same manner as in \cite{alasseur2024deep}, leading to a more flexible and numerically tractable scheme.

\vspace{1em}

To approximate the solution numerically, let $\pi = \{t_0, \dots, t_M\}$ be a uniform partition of $[0,T]$ with $\Delta t = T/M$. Define $\Delta N_i \sim \text{Poisson}(\lambda \Delta t)$ as the number of jumps on $[t_i, t_{i+1}]$, and $\Delta J_l^i$ as the corresponding jump sizes, which are independent and distributed according to the L\' evy measure $\nu$.

The forward SDE is discretized following standard Euler-Maruyama with jumps scheme as in \cite{bruti2007strong}. Specifically, we approximate 
\begin{align*}
X_{i+1}^{\pi} = X_i^{\pi} + \bar{b}(X_i^{\pi}) \Delta t + \sum_{l=1}^{\Delta N_i} \beta(X_i^{\pi}, \Delta J_l^i),
\end{align*}
where $\bar{b}(x) = b(x) - \int \beta(x,z) \nu(dz)$.

The BSDE with jumps is also approximated by a standard Euler-Maruyama scheme 
\begin{align*}
Y_i^{\pi} \approx Y_{i+1}^{\pi} + f(t_i, X_i^{\pi}, Y_i^{\pi}) \Delta t - \left( u(t_i, X_i^{\pi} + \sum_l \beta(X_i^{\pi}, \Delta J_l^i)) - u(t_i, X_i^{\pi}) \right).
\end{align*}

We train two neural networks:
\begin{itemize}
    \item Value network $\mathcal{N}_n^Y$: It takes as input current time $t_n$ and state $X_n^\pi$ and approximates $Y_n^{\pi}$.
    \item Jump network $\mathcal{N}_n^U$: It takes  as input current time $t_n$, state $X_n^\pi$ and jump coefficient $\beta(X_i^\pi, \Delta J_l^i)$ and approximates $U_n^{\pi}$.
\end{itemize}

The overall architecture is trained using stochastic gradient descent based on the temporal residuals of the BSDE system.

\begin{algorithm}[H]
\caption{DeepBackwardBSDE-JumpSolver}
\label{algo:jumpFBSDE}
\begin{algorithmic}[1]
\State \textbf{Input:} Time grid $\pi$, batch size $B$, number of epochs $E$, learning rate $\eta$
\For{$n = M-1$ to $0$}
    \For{epoch $e = 1$ to $E$}
        \For{$j = 1$ to $B$} \Comment{Batch loop}
            \State Initialize $X_0^{\pi,j} = x_0$
            \For{$i = 0$ to $n$}
                \State Sample $\Delta N_{i+1}^j \sim \text{Poisson}(\lambda \Delta t)$
                \State Sample jumps $(\Delta J_l^{i+1})_{l=1}^{\Delta N_{i+1}^j} \sim \nu/\lambda$
                \State $X_{i+1}^{\pi,j} \gets X_i^{\pi,j} + b(t_i, X_i^{\pi,j}) \Delta t + \sum_{l=1}^{\Delta N_{i+1}^j} \beta(t_i, X_i^{\pi,j}, \Delta J_l^{i+1})$
            \EndFor
            \State $Y_n^j \gets \mathcal{N}_n^Y(t_n, X_n^{\pi,j}; \theta_n)$
            \State $U_n^j \gets \mathcal{N}_n^U\left(t_n, X_n^{\pi,j}, \sum_{l=1}^{\Delta N_n^j} \beta(t_n, X_n^{\pi,j}, \Delta J_l^n) ; \theta_n \right)$
            \If{$n = M-1$}
                \State $Y_{M}^j \gets g(X_{M}^{\pi,j})$
            \Else
                \State $Y_{n+1}^j \gets \mathcal{N}_{n+1}^Y(t_{n+1}, X_{n+1}^{\pi,j}; \theta_n^*)$
            \EndIf
        \EndFor

        \For{$k = 1$ to $A$} \Comment{Auxiliary batch for conditional expectation}
            \State Sample $\Delta \Bar{N}_n^k$, $\Delta \Bar{J}_l^{n,k}$ analogously
            \State $W_n^k \gets \mathcal{N}_n^U\left(t_n, X_n^{\pi,j}, \sum_{l=1}^{\Delta \bar{N}_n^k} \beta(t_n, X_n^{\pi,j}, \Delta \bar{J}_l^{n,k}); \theta \right)$
        \EndFor

        \State Compute loss:
        \[
        \ell(\theta_n) = \frac{1}{B} \sum_{j=1}^B \left|Y_{n+1}^j - \left(Y_n^j - f(t_n, X_n^{\pi,j}, Y_n^j) \Delta t + U_n^j - \frac{1}{A} \sum_{k=1}^A W_n^k \right)\right|^2
        \]
        \State Update network parameters using SGD: \quad \( \theta_n \leftarrow \theta_n - \eta \nabla_{\theta_n} \ell(\theta_n) \)
    \EndFor
    \State 
    \( \theta_n^* \leftarrow  \theta_n \)
\EndFor 
\State \Return \( (\mathcal{N}^Y_n, \mathcal{N}^U_n) \) for all \( n = 0, \dots, M \)
\end{algorithmic}
\end{algorithm}

\subsection{Optimal Control Selection via Deep BSDE Solver}
\label{sec:control}
We now revisit the specialized two-dimensional model introduced in Section \ref{sec:2dim_model}, focusing on the optimal selection of the threshold control parameter \( A \in \mathcal{A} \subset \bR_+ \), which governs installation decisions. The state variables \( V(t), D(t), C^A_R(t) \) evolve according to threshold-based jump dynamics as discussed previously, and the associated cost functional is:
\begin{equation*}
\begin{aligned}
J^A(0,v,d,c) = \mathbb{E} \bigg[ \kappa C^A_R(T) + \int_0^T e^{-rs}\left(D(s) - V(s) C^A_R(s)\right)^+ ds \\ \bigg|\, V(0) = v, D(0) = d, C^A_R(0) = c \bigg],
\end{aligned}
\end{equation*}

\paragraph{FBSDE Formulation.}
The control problem admits a probabilistic representation in terms of a BSDE with jumps. Define the state vector \( X^A(t) = (V(t), D(t), C^A_R(t)) \), which evolves according to a jump-diffusion process with drift and jump coefficients determined by the capacity and demand dynamics and the threshold rule:
\begin{equation}
\label{eq:FBSDE-optimal}
\begin{cases}
-dY(t) =e^{-rt} \left( D(t) - V(t) C^A_R(t) \right)^+ dt - \int_{\mathbb{R}_+} U_1(t,z) \, \widetilde{N}_1(dt,dz) - \int_{\mathbb{R}_+} U_2(t,z) \, \widetilde{N}_2(dt,dz), \\
Y(T) = \kappa C^A_R(T).
\end{cases}
\end{equation}
The terms \( U_i(t,z) \) represent the BSDE's sensitivity to jumps in the capacity factor process, where jumps trigger control actions according to:
\begin{equation*}
\Delta C^A_R(t) = (A - V(t))^+ \Delta V(t).
\end{equation*}

\paragraph{Control Selection Strategy.}
To identify the optimal control policy \( A^* \), we approximate the admissible set $\cA$ by a finite grid \( \mathbb{A} = \{A_1,  A_2 \dots, A_{N_A}\}\) and evaluate the expected cost \( Y^A(0) \) for each candidate \( A \) by solving the BSDE \eqref{eq:FBSDE-optimal} using the neural solver introduced in Section \ref{sec:DeepBSDE}. The threshold that minimizes this value defines the optimal installation rule.

\begin{remark}
    While we focus here on the threshold-based rule \eqref{eq:threshold-1d}, other threshold-based policies could also be considered. In practice, one can explore different structures or combinations of thresholds and select the one that minimizes the expected cost. The one strategy proposed turned out to be the best performing one. 
\end{remark}

The {DeepBSDE-ControlSelector} algorithm is designed to identify the optimal threshold parameter \( A^* \) for renewable capacity installation. This is done by minimizing the expected total cost functional associated with the BSDE system \eqref{eq:FBSDE-optimal}, which captures both terminal installation costs and operational mismatches between demand and renewable supply.

\begin{remark}
The novelty of this approach lies in its use of a BSDE-based numerical solver to address a control selection problem. Rather than solving the HJB equation directly as is typical in control theory, we leverage the BSDE formulation of the value function to evaluate the cost associated with each candidate control. This effectively bypasses the need for solving a high-dimensional HJB PIDE and instead relies on deep learning-based BSDE methods to perform control selection in a tractable and flexible manner.
\end{remark}

The algorithm proceeds as follows:

\begin{algorithm}[H]
\caption{DeepBSDE-ControlSelector}
\label{algo:control-selection}
\begin{algorithmic}[1]
\State \textbf{Input:}
\begin{itemize}
    \item Bounds \( A_{\min}, A_{\max} \in \bR_+  \)
    \item Number of candidate thresholds \( N_A \in \mathbb{N} \)
\end{itemize}

\State \textbf{Step 1: Construct candidate grid}
\State Generate a uniform grid of thresholds:
\[
\mathbb{A} = \{ A_1, A_2, \dots, A_{N_A} \}, \quad \text{where } A_1 = A_{\min}, A_{N_A} = A_{\max}
\]

\State \textbf{Step 2: Evaluate cost for each candidate}
\For{each \( A \in \mathbb{A} \)}
    \State Apply DeepBackwardBSDE-JumpSolver (Algorithm \ref{algo:jumpFBSDE}) under the forward process $X^A(t)$
    \State Store the pair \( (A, Y^A(0)) \)
\EndFor

\State \textbf{Step 3: Select optimal control}
\State Identify the control that minimizes the expected cost:
\[
A^* = \arg\min_{A \in \mathbb{A}} Y^A(0), \quad Y_0^* = \min_{A \in \mathbb{A}} Y^A(0)
\]

\State \textbf{Output:} Optimal threshold \( A^* \) and associated minimal value \( Y_0^* \)
\end{algorithmic}
\end{algorithm}

\section{Numerical Implementation}
\label{sec:numerics}

In this section, we present the results obtained by applying the DeepBSDE-ControlSelector (Algorithm~\ref{algo:control-selection}) to solve the optimal control problem introduced in Section~\ref{sec:2dim_model}, focusing on the particular two-dimensional case. The method involves repeatedly solving the BSDE \eqref{eq:FBSDE-optimal} through DeepBackwardBSDE-JumpsSolver (Algorithm~\ref{algo:jumpFBSDE}) for various fixed and constant values of the control parameter \( A \in \mathbb{A} \).

\subsection{Neural Network Architecture}
To approximate the solution, we employ fully connected feedforward neural networks. Each network consists of an input layer, \( L = 2 \) hidden layers, each containing \( 100 \) neurons and one output layer. All hidden layers use the \textit{tanh} activation function. Network weights at each timestep are initialized using optimal weights from the previous timestep. 
The optimizer used is Adam, with a learning rate set to \( 10^{-4} \). We employ a batch size of \( 10 \), while we use \( 5000 \) samples to estimate the compensator. The time horizon is discretized into \( M = 50 \) time steps. Training is carried out for \( 4000 \) epochs at timestep $M-1$ and for \(200\) epochs for all the other timesteps. All neural networks are implemented in PyTorch.

\subsection{Numerical Results}
\label{sec:num_Results}
\begin{figure}[h!]
    \centering

    \begin{subfigure}[b]{0.32\linewidth}
        \includegraphics[width=\linewidth]{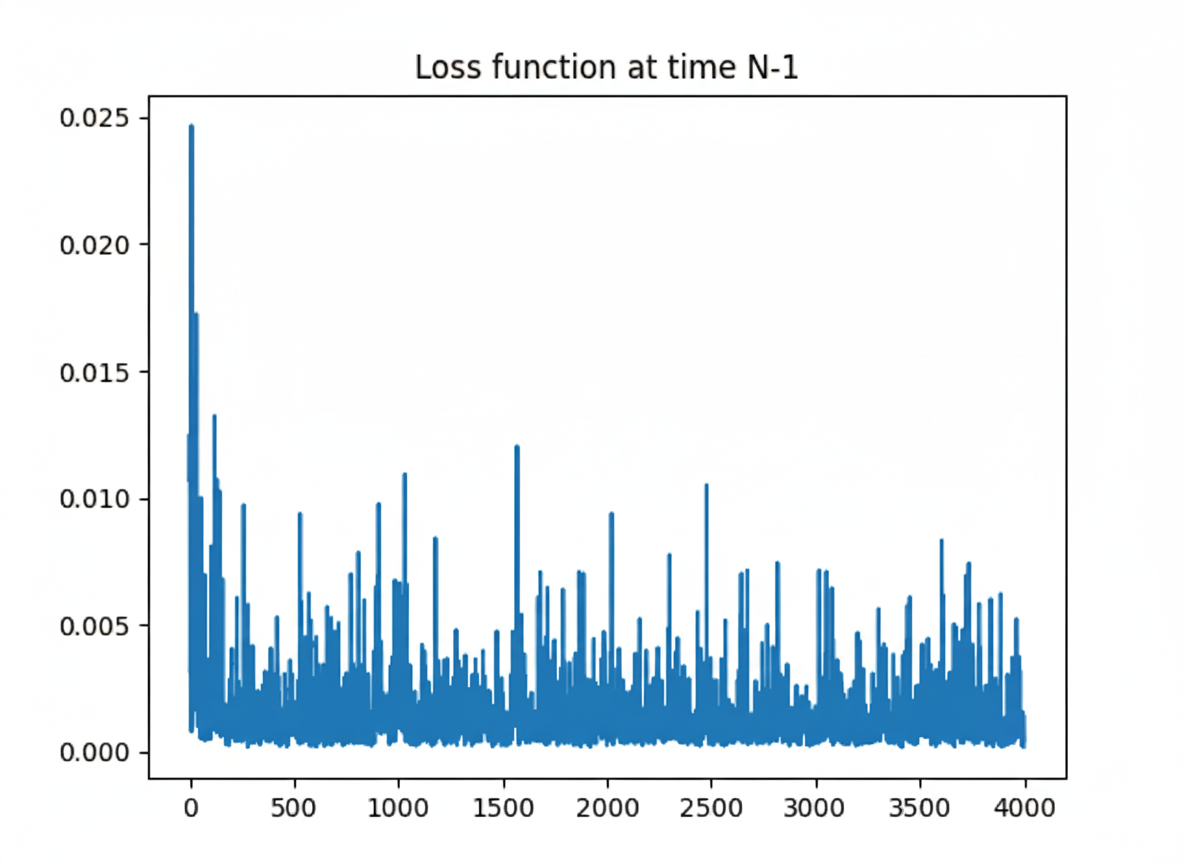}
        \label{fig:loss1}
    \end{subfigure}
    \hfill
    \begin{subfigure}[b]{0.32\linewidth}
        \includegraphics[width=\linewidth]{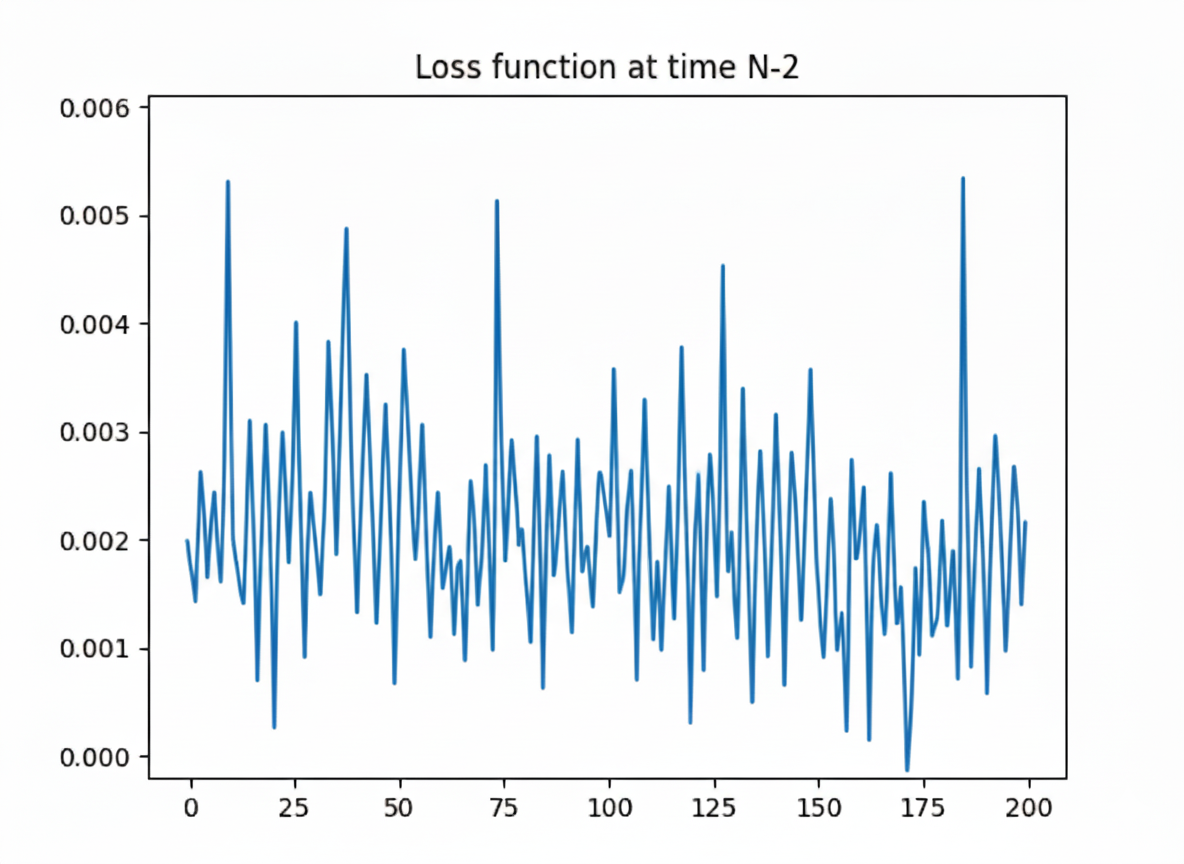}
        \label{fig:loss2}
    \end{subfigure}
    \hfill
    \begin{subfigure}[b]{0.32\linewidth}
        \includegraphics[width=\linewidth]{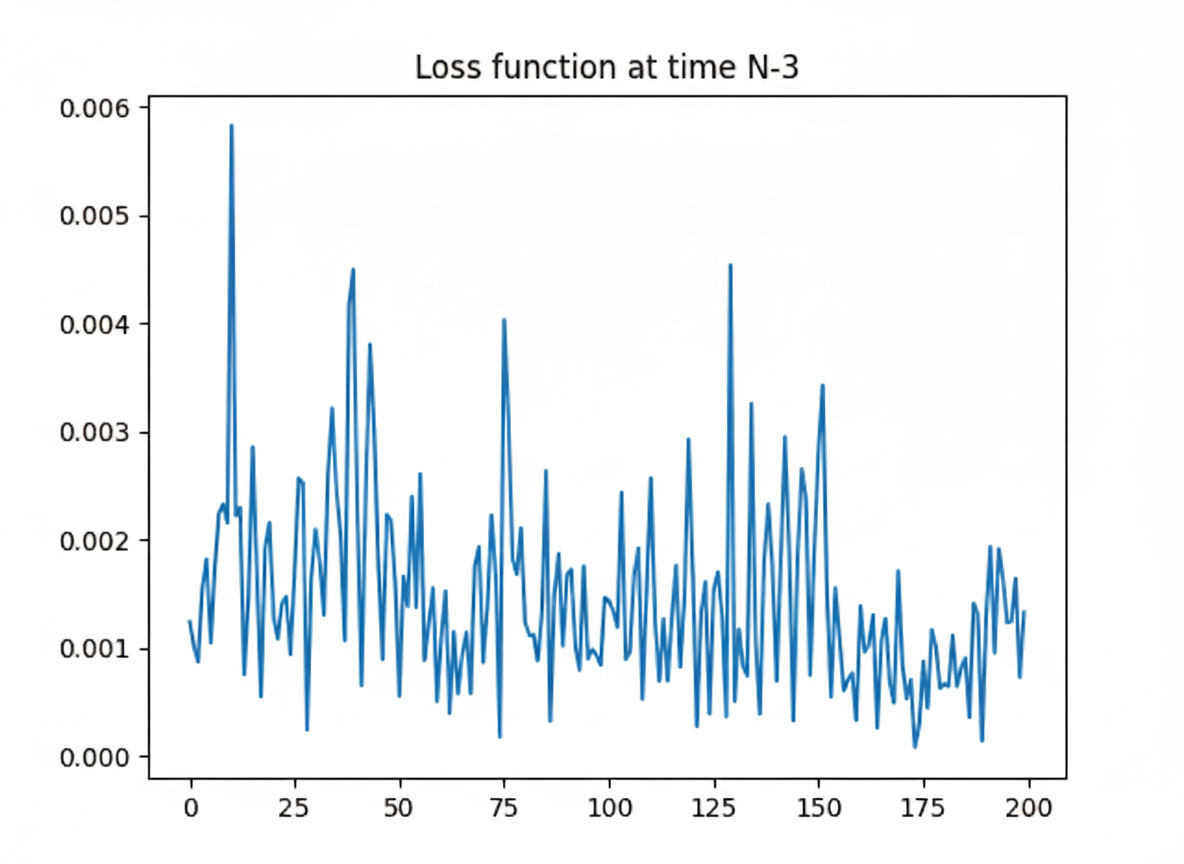}
        \label{fig:loss4}
    \end{subfigure}
\caption{Convergence of loss function at different timesteps}
 \label{fig:losses}
\end{figure}

We begin by outlining the model configuration used in our numerical experiments. The simulation involves the forward processes \( V(t) \), \( D(t) \), and \( C_R^A(t) \), specified in equations \eqref{eq:V_din}, \eqref{eq:D_din}, and \eqref{eq:threshold-1d}. These processes, along with the cost functional defined in \eqref{eq:cost-1d}, form the basis of the optimal control problem. The corresponding model parameters are summarized in Table~\ref{tab:param}. 
Note that functions $p(t)$ and $s(t)$ are assumed to be constants.
\small
\begin{table}[H]
\renewcommand{\arraystretch}{1.5}
\centering
\begin{tabular}{|c|c|c|c|c|c|c|c|c|c|c|}
\hline
\textbf{$T$} & \textbf{$x^0$} & \textbf{$\lambda$} & \textbf{$m$} & \textbf{$\sigma$} & \textbf{$\xi$} & \textbf{$p$} & \textbf{$s$} & \textbf{$r$} & \textbf{$\kappa$} \\
\hline
1 & [0.4, 0.7, 0.0] & [5.0, 5.0] & [0.5, 1.0] &
$\begin{bmatrix}
0.2 & 0.2 \\
0.0 & 0.05
\end{bmatrix}$ &
[0.2, 0.2] & 0.7 & 1.0 & 0.4 & 0.1 \\
\hline
\end{tabular}
\caption{Model parameters}
\label{tab:param}
\end{table}
In Figure~\ref{fig:losses}, we report the training loss associated with the last three time steps of the DeepBackwardBSDE-JumpsSolver. Since the algorithm operates locally in time, we optimize a separate neural network at each time step, which produces its own loss trajectory. The plots shown correspond to the final three time steps; the losses at earlier times exhibit similar behavior and are omitted for brevity.

\begin{figure}[H]
    \centering
    \includegraphics[width=0.5\linewidth]{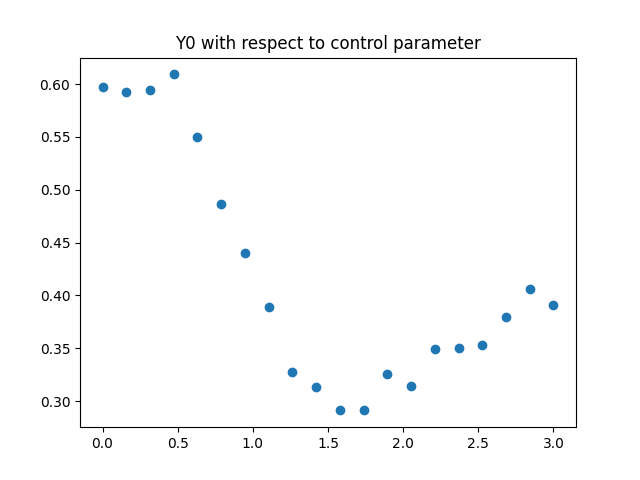}
    \caption{A scatter plot of $Y(0)$ values corresponding to different values of $A$}
    \label{fig:scatter}
\end{figure}

To determine the optimal value of the threshold parameter \( A \), we run Algorithm~\ref{algo:control-selection} for $A_{\min} = 0$ and $A_{\max} = 3$. Specifically, we evaluate the estimated value of \( Y(0) \) for $N_A = 20$ equidistant choices of \( A \). The resulting estimates are summarized in Figure~\ref{fig:scatter}, which provides a visual overview of how the initial value of the backward process depends on the choice of \( A \).

From Figure~\ref{fig:scatter}, we observe that the optimal threshold level appears to be at \( A \approx 1.58 \), corresponding estimated value of \( Y(0) \approx 0.29 \).

\begin{figure}[H]
    \centering

    \begin{subfigure}[b]{0.48\linewidth}
        \label{fig: state}
        \includegraphics[width=\linewidth]{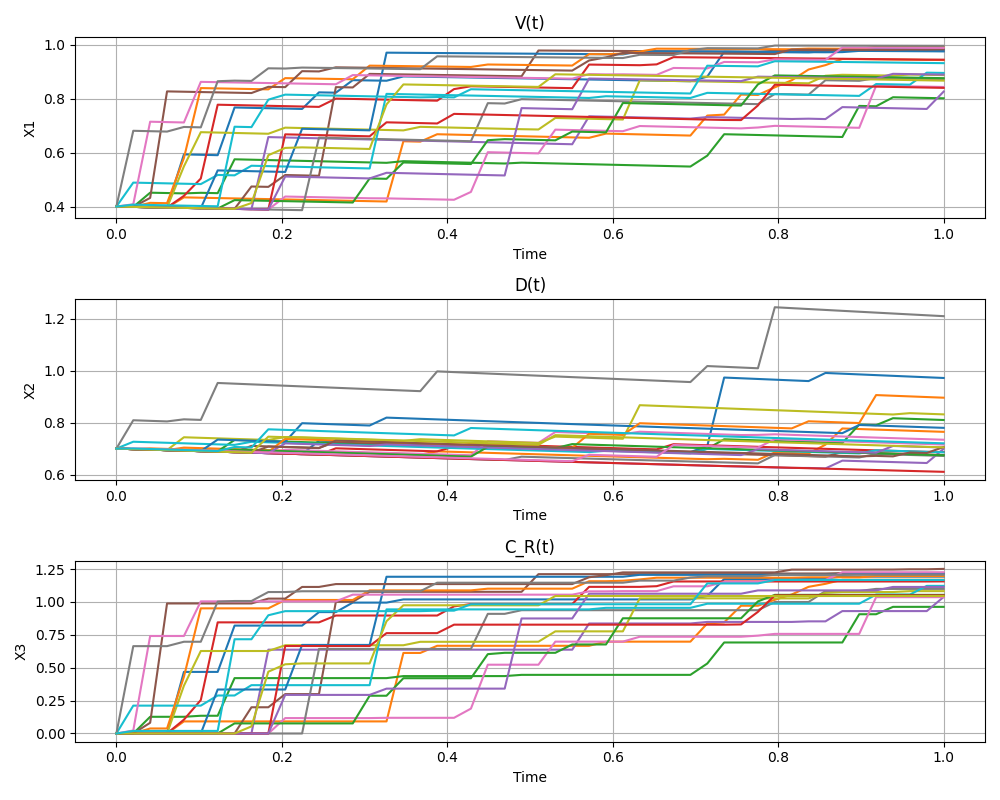}
    \end{subfigure}
    \hfill
    \begin{subfigure}[b]{0.48\linewidth}
        \label{fig:Yt}
        \includegraphics[width=\linewidth]{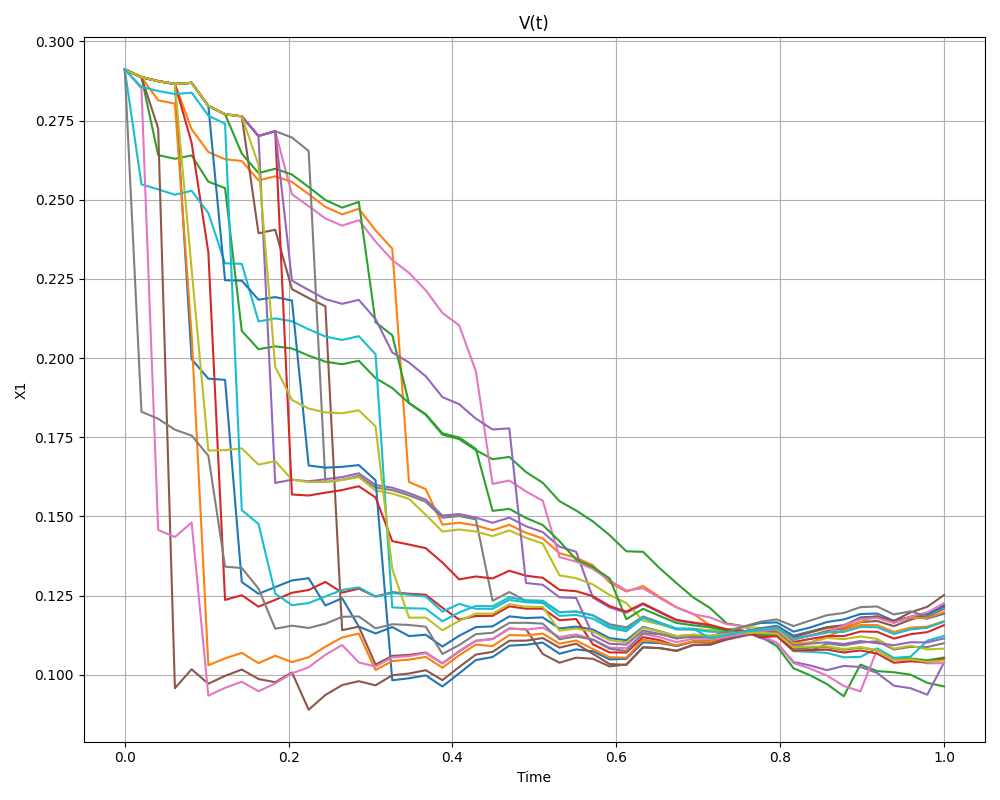}
    \end{subfigure}
\caption{Left: sample paths of the forward processes 
$V(t), D(t), C_R^A(t)$ Right: corresponding realizations of the backward component 
$Y(t)$ obtained from the BSDE solution.}
\label{fig:realizations}
\end{figure}

In Figure~\ref{fig:realizations}, we illustrate the outcome of the forward-backward simulation when the control is fixed to the optimal value \( A \approx 1.58 \). The left subfigure shows sample trajectories of the three-dimensional forward process \( (V(t), D(t), C_R^A(t)) \), while the right one reports the corresponding realisations of the backward component \( Y \) obtained from the BSDE solution. This figure helps us to better visualise the temporal evolution of the state process and the associated value function under the learned policy. Notice that installations of renewable capacities are happening multiple times throughout the time horizon.

\section{Deep Control Approach}
\label{sec:deepcontrol}
In this section, we present an alternative numerical strategy for solving the renewable installation control problem by leveraging global deep learning, inspired by the method introduced in \cite{ han-control}. 

Unlike the threshold-based control rule of Section~\ref{sec:control-model}, which triggers discrete installations only when the capacity factor drops below a predetermined level, we now seek to directly learn a state-dependent feedback control. However, the installation policy remains driven by jumps in the capacity factor. This feedback strategy aims to determine, at any time \( t \in [0,T] \), the optimal intensity of renewable capacity installation based on the full system state.

We consider the same state variables as in the specialized case of Section \ref{sec:2dim_model}: the renewable capacity factor \( V(t) \), the demand process \( D(t) \), and the installed capacity \( C_R^A(t) \). The evolution of the system is governed by the pure-jump dynamics:
\[
X(t) = (V(t), D(t), C_R^A(t))^\top \in \mathbb{R}_+^3,
\]
driven by two independent L\' evy measures \( N_1(t) \) and \( N_2(t) \) with intensities \( \lambda_1 \), \( \lambda_2 \) and jump distributions \( \nu_1 \), \( \nu_2 \), respectively.

We now generalize the installation rule by letting the update to capacity be determined by a feedback control \( A(t,x) \) at time \( t \) and state \( x = (v,d,c) \). Processes \(V(t)\) and \(D(t)\) follow the dynamics specified in \eqref{eq:V_din} and \eqref{eq:D_din}. The control acts only at jump times of \( V(t) \), via:
\begin{equation*}
\begin{aligned}
dC_R^A(t) &= A_1(t, X(t^-))d\widetilde{N}_1(dt,dz) +A_2(t, X(t^-))d\widetilde{N}_2(dt,dz).
\end{aligned}
\end{equation*}

 Notice that we introduce two separate controls, $A_1$ and $A_2$, corresponding to the two jump sources $\widetilde{N}_1$ and $\widetilde{N}_2$. This allows us to distinguish between capacity adjustments triggered solely by jumps in $V$ (via $A_1$) and those also influenced by jumps in $D$ (via $A_2$), in accordance with the dynamics specified in \eqref{eq:V_din} and \eqref{eq:D_din}. The control amplitude \( A_i(t, x) \in \mathbb{R}_+ \) for $i=1,2$, is now a learned function approximated by a neural network \( \mathcal{A}(t,x;\theta) \).

All stochastic integrals are interpreted in the pure jump setting, and \( V(t) \) is given explicitly as a nonlinear function of an underlying OU process (as described in Section \ref{sec:2dim_model}).


The performance of the policy \( A \) is evaluated via the same cost functional $J$ defined in \eqref{eq:cost-1d}.
 The goal is to find a feedback control \( A^*(t,x) \) that minimizes this cost. Unlike the previous BSDE-based method, this approach learns the control directly by minimizing an empirical approximation of \( J \). Our approach also differs from the one recently proposed in \cite{control-siam}, where a deep learning estimation of the value function is performed in combination with the stochastic Pontryagin maximum principle. 

\subsection{Learning the Feedback Policy via Neural Networks}

\begin{figure}[H]
    \centering
    \includegraphics[width=0.9\textwidth]{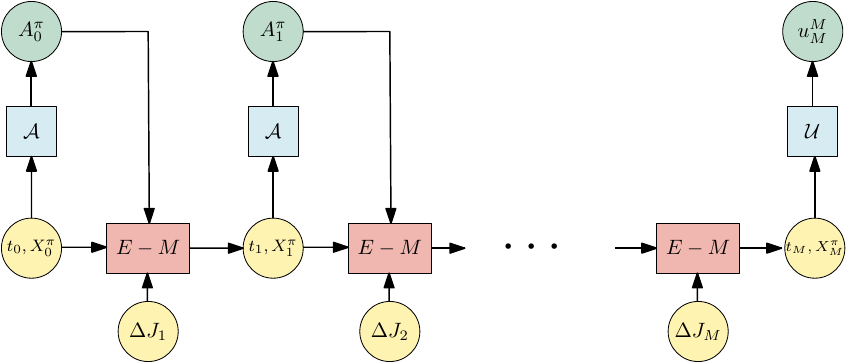}
    \caption{Schematic representation of the neural network policy \( \mathcal{A}(t, x; \theta) \) interacting with the system. Here, $E-M$ stands for Euler-Maruyama discretization. }
    \label{fig:nn-policy-diagram}
\end{figure}

We approximate the optimal control \( A^*(t,x) \) by a feedforward neural network \( \mathcal{A}(t,x;\theta) \) with parameters \( \theta \). The network takes time \( t \) and state \( x = (v,d,c) \) as input and outputs a non-negative scalar control action. The training procedure involves simulating many trajectories of the system under \( \mathcal{A} \) and minimizing the average cost across those trajectories. Figure~\ref{fig:nn-policy-diagram} illustrates how the neural network policy interacts with the system over time. At each discrete time step, the current state is passed into the network, which outputs a control action. This influences the next state and cost, and the process repeats across the trajectory.

We summarize the implementation in Algorithm \ref{algo:control}.

\begin{algorithm}[H]
\caption{DeepControlSolver}
\label{algo:control}
\begin{algorithmic}[1]
\For{each epoch}
    \State Initialize all paths: $X_0^{\pi,j} = x_0$ for $j = 1, \dots, B$
    \For{$n = 0$ to $M-1$}
        \For{$j = 1$ to $B$}
            \State Evaluate feedback control: $A_n^{\pi,j} = \mathcal{A}(t_n, X_n^{\pi,j}; \theta)$
            \State Sample Poisson increment: $\Delta N_{n+1}^j \sim \text{Poisson}(\lambda \Delta t)$
            \State Sample jump sizes: $\{ \Delta J_l^{n+1} \}_{l=1}^{\Delta N_{n+1}^j} \sim \nu/\lambda$
            \State Compute total jump: $\Delta V_{n+1}^j := \sum_l \text{JumpImpact}(X_n^{\pi,j}, \Delta J_l^{n+1})$
            \State Update state:
            \[
            X_{n+1}^{\pi,j} \gets X_n^{\pi,j} + b(t_n, X_n^{\pi,j}) \Delta t + \beta(t_n, X_n^{\pi,j}, A_n^{\pi,j}, \Delta V_{n+1}^j)
            \]
        \EndFor
    \EndFor
    \State Compute empirical loss:
    \[
    \ell(\theta) = \frac{1}{B} \sum_{j=1}^B \left[ \Delta t \sum_{n=0}^{M-1} e^{-r n \Delta t }(X_{2,n}^{\pi,j} - X_{1,n}^{\pi,j} X_{3,n}^{\pi,j})^+ + \kappa \cdot X_{3,M}^{\pi,j} \right]
    \]
    \State Update weights: \( \theta \gets \theta - \eta \nabla_\theta \ell(\theta) \)
\EndFor
\State \Return Trained policy \( \mathcal{A}(t,x;\theta) \)
\end{algorithmic}
\end{algorithm}

\subsection{Numerical Results}

The network consists of an input layer, \( L = 2 \) hidden layers, each containing \( 256 \) neurons and one output layer. All hidden layers use the \textit{ReLU} activation function. Network weights at each timestep are initialized using the Kaiming normal initialization \cite{he2015delving}. 
The optimizer used is Adam, with a learning rate set to \( 10^{-4} \). We employ a batch size of \( 2000 \). The time horizon is discretized into \( M = 50 \) time steps. Training is carried out for \( 50 \) epochs. All neural networks are implemented in PyTorch.

\begin{figure}[h!]
    \centering

    \begin{subfigure}[b]{\linewidth}
        \centering
        \includegraphics[width=0.7\linewidth]{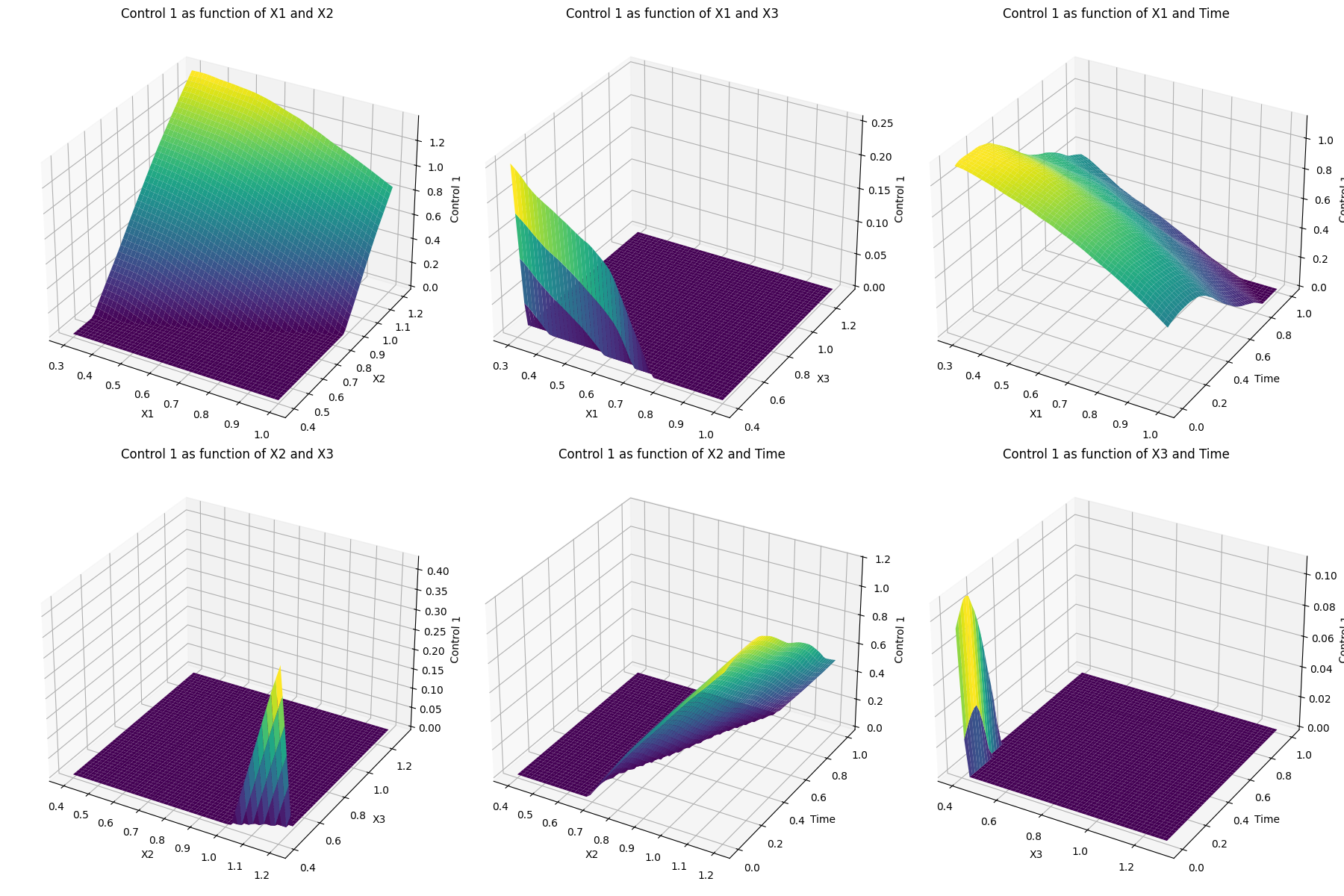}
        \caption{First component}
        \label{fig:control-1}
    \end{subfigure}

    \vspace{0.5em}

    \begin{subfigure}[b]{\linewidth}
        \centering
        \includegraphics[width=0.7\linewidth]{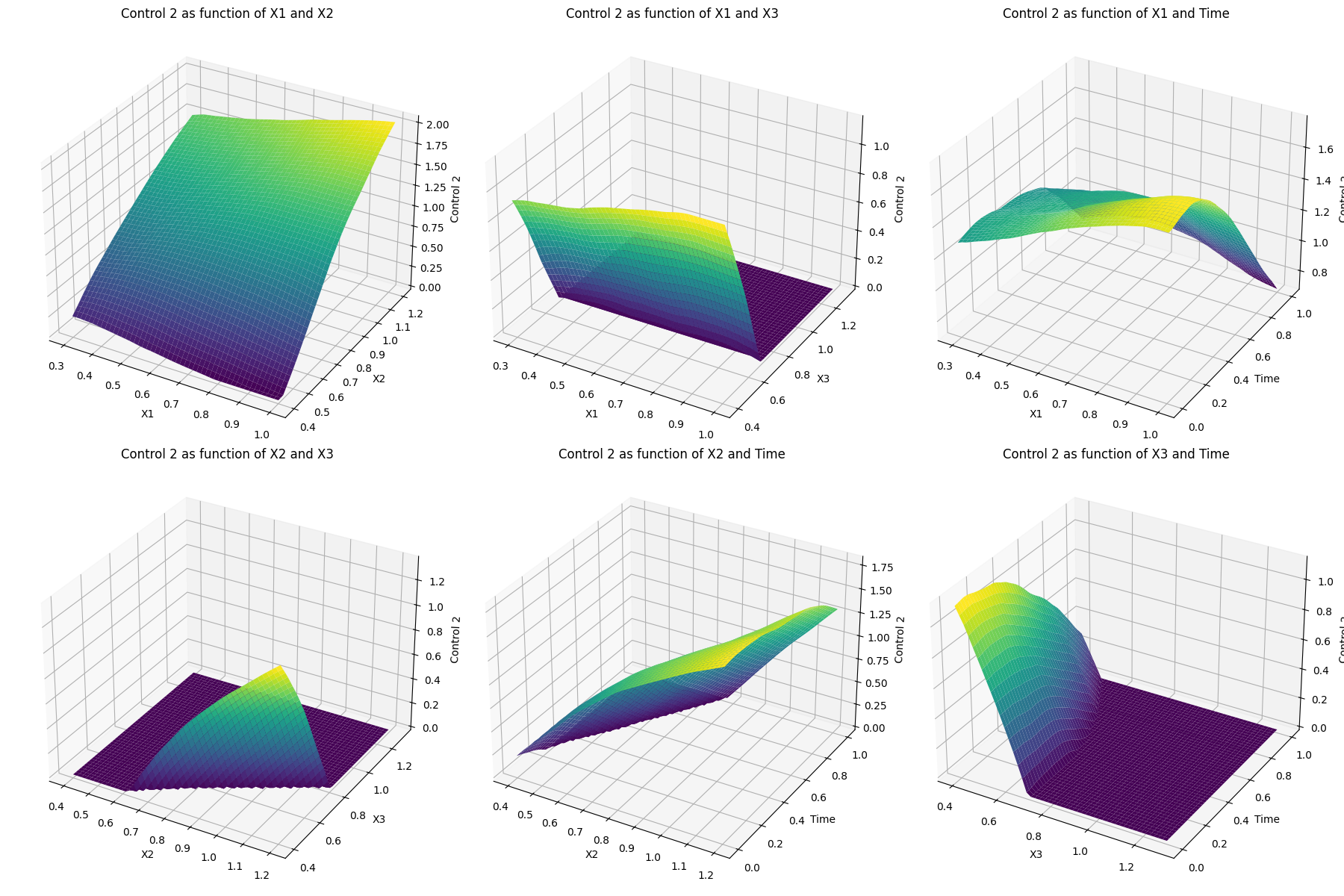}
        \caption{Second component}
        \label{fig:control-2}
    \end{subfigure}

    \caption{Control $A$ as a function of its arguments}
    \label{fig:control-surface}
\end{figure}

In Figure \ref{fig:control-surface}, we provide numerical comparisons and surface plots showing how the learned feedback control \( A(t,x) \) varies with respect to state variables and time, as well as convergence behavior of the loss function.

\begin{figure}
    \centering
    \includegraphics[width=0.5\linewidth]{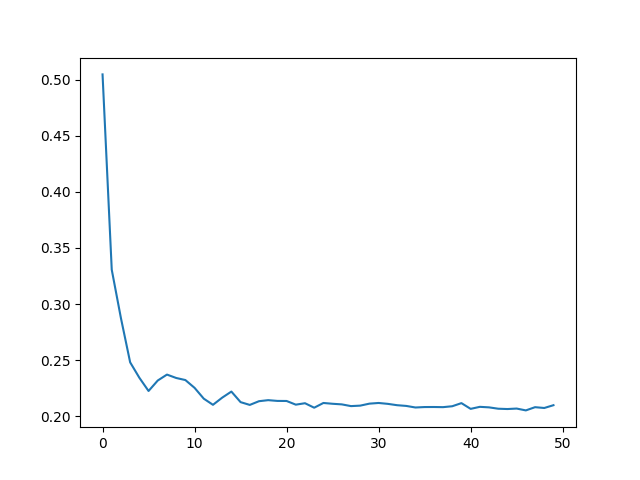}
    \caption{Convergence of the loss function}
    \label{fig:control-loss}
\end{figure}

In Figure \ref{fig:control-loss} we present some results obtained through the above algorithm, in particular, we present the convergence of the loss function, which corresponds to the performance functional $J$ and surface plots of the feedback control $A$ with respect to two of its four arguments. 

\begin{figure}[h!]
    \centering

    \begin{subfigure}[b]{0.48\linewidth}
        \includegraphics[width=\linewidth]{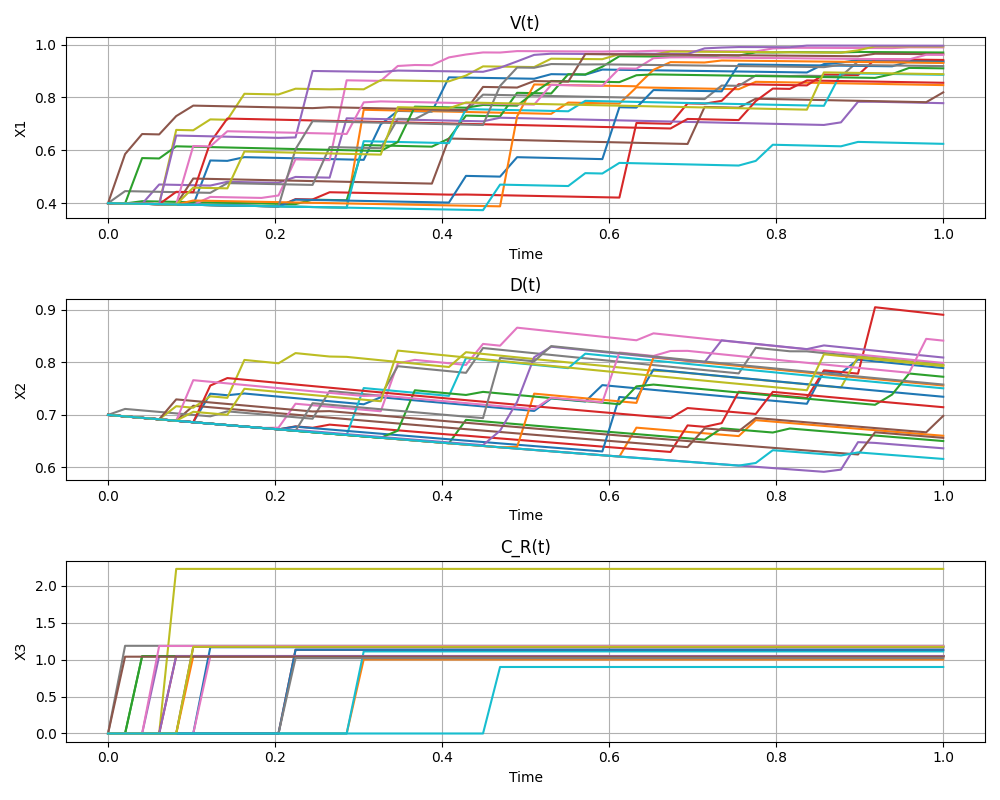}
        \label{fig:control-state}
        \caption{State}
    \end{subfigure}
    \hfill
    \begin{subfigure}[b]{0.48\linewidth}
        \includegraphics[width=\linewidth]{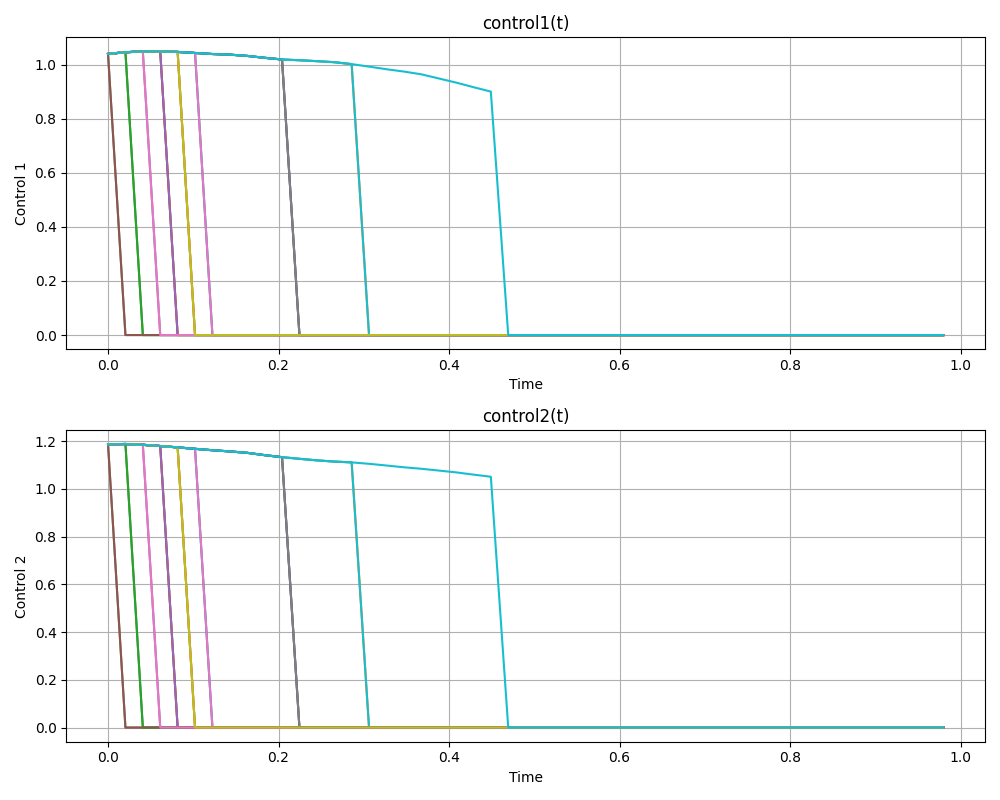}
        \label{fig:control-control}
        \caption{Control}
    \end{subfigure}
\caption{Realizations}
\label{fig:control-dynamics}
\end{figure}

At last, let us also present in Figure \ref{fig:control-dynamics}, 20 realizations of state and control dynamics similar to the ones presented in Figure \ref{fig:realizations}. We can see that under this approach, where control is also time-dependent, we reduce the level of action when we get closer to the terminal time. In Figure \ref{fig:control-surface}, we observe how the control is constant in the sum of the first two state components while it decays with time. Furthermore, in each scenario, installation is performed at a single time, which is consistent with the findings in \cite{agram2025installation}. This flexibility, together with dependence on the state, which is evident in Figure \ref{fig:control-surface}, is the reason why we outperform the constant threshold approach from the previous section. Indeed, in Figure \ref{fig:control-loss} we can see we are able to achieve a value of less than 0.21 compared to the minimal value of 0.29 as seen in Figure \ref{fig:scatter}.

\newpage

\section{Conclusion}
\label{sec:conclusion}

In this paper, we developed a data-driven stochastic control framework for the installation of renewable energy capacity under uncertainty. The system dynamics are modeled by a multidimensional OU process with jump components, capturing abrupt fluctuations in both renewable availability and demand.

We considered two solution strategies. The first is a structured threshold-based policy, in which investment is triggered when the stochastic capacity factor drops below a prescribed level. This leads to a nonlinear PIDE, which we reformulated via a Feynman-Kac representation as a BSDE with jumps. We then proposed a neural solver extending the backward in time deep BSDE method of \cite{hure2020deep} to account for discontinuous drivers. This approach, implemented with a dual network architecture, enables tractable evaluation of interpretable rule based strategies in high dimensional settings.

The second approach, the \emph{DeepControlSolver}, relaxes the structural constraints of the threshold policy by directly parameterizing the feedback control law as a neural network and optimizing it through stochastic gradient descent. This formulation bypasses the PIDE-BSDE link, allows for fully state and time dependent decisions, and often achieves lower cost values in simulations.

Our numerical experiments indicate that, under the chosen parameter regime, the \emph{DeepControlSolver} identifies the mathematically optimal strategy, consisting of a single installation over the planning horizon, consistently with the findings in \cite{agram2025installation}. On the other hand, implementing the threshold-based policy of \emph{DeepBSDE-ControlSelector} may appear more practical in real-world contexts: multiple interventions allow investments and resources to be spread over time, reflecting common practice, and the policy provides a transparent and interpretable rule specifying what to install and when.

In conclusion, while the \emph{DeepControlSolver} achieves better performance, the threshold-based approach of \emph{DeepBSDE-ControlSelector} remains attractive due to its interpretability and ease of implementation. The observed difference in performance can be viewed as the "cost" associated with following a more interpretable policy. Together, these methods offer complementary tools for the analysis and design of renewable capacity expansion strategies under stochastic jump dynamics.


\bibliographystyle{plain}
\bibliography{references}
\end{document}